# Precision Glass Thermoforming Assisted by Neural Networks

Yuzhou Zhang[1], Mohan Hua[1], Jinan Liu[2], Haihui Ruan[1, 3*]

1. Department of Mechanical Engineering, The Hong Kong Polytechnic University, Hung Hom, Hong Kong
2. Biel Crystal (HK) Manufactory Ltd., Hong Kong, Hong Kong
3. PolyU-Daya Bay Technology and Innovation Research Institute, Huizhou, Guangdong, China

## ABSTRACT

Many glass products require thermoformed geometry with high precision. However, the traditional approach of developing a thermoforming process through trials and errors can cause large waste of time and resources and often end up with unsuccessfulness. Hence, there is a need to develop an efficient predictive model, replacing the costly simulations or experiments, to assist the design of precision glass thermoforming. In this work, we report a surrogate model, based on a dimensionless back-propagation neural network (BPNN), that can adequately predict the form errors and thus compensate for these errors in mold design using geometric features and process parameters as inputs. Our trials with simulation and industrial data indicate that the surrogate model can predict forming errors with adequate accuracy. Although perception errors (mold designers' decisions) and mold fabrication errors make the industrial training data less reliable than simulation data, our preliminary training and testing results still achieved a reasonable consistency with industrial data, suggesting that the surrogate models are directly implementable in the glass-manufacturing industry.

[*] Corresponding author: Tel.: + 852 2766 6648, Fax: +852 2365 4703, E-mail: haihui.ruan@polyu.edu.hk

# 1. Introduction

Glass, here referring to various optically transparent oxides, is an important part of modern life and sciences due to its high hardness and chemical inertness and its ability to deflect light paths with little loss in intensity. The early scientific giants, such as Galileo, Descartes, Newton, and Fraunhofer, all devoted significant effort to producing glass lenses [1, 2]. Nowadays, the continuous research in glass is propelled by the development of novel technologies and the advocation for sustainable development. This is represented by the emergence of the fifth-generation (5G) wireless communication technology and the increasing applications based on artificial intelligence (AI) technologies, demanding many more optical products than before and their quick updating rates [3, 4]. Today, the most representative application of glass is probably glass covers for smartphones and AR/VR devices. These applications are due to the good processability of glass through thermoforming and the adequate wear and impact resistance of glass through thermochemical strengthening. However, as the shapes of these covers become increasingly complex, which generally require precisely curved profiles with varied geometrical features and thicknesses, it poses a basic question to the glass manufacturing industry: how to make these curved glass products accurately, efficiently, and with high yield.

The traditional methods to fabricate precision glass products are through grinding, polishing, and lapping, which are very time-consuming and expensive. For example, single-point diamond turning is an effective method to fabricate complex glass profiles with high precision, but it requires several or tens of hours to produce one piece [5]. Therefore, lens-makers have switched to the precision glass molding (PGM) technology to form glass products at high temperatures where the viscosity of glass is in the range of $10^7$ to $10^8$ Pa·s. Since the earlier time of this century, PGM has been adopted by many optical manufacturers to fabricate aspherical lenses [6, 7] and more complex optical components (e.g., micro-lens arrays) [8, 9]. The high precision in PGM is achieved through the principle of error compensation, that is the form errors in the final glass product are compensated by adding error-compensation profiles in mold designs.

Designing precision molds with forming error compensation is ubiquitously used in many forming processes. For example, in metal forming, Gan and Wagoner [10] proposed a method named 'displacement adjustment' (DA) to design the mold profiles to compensate for form errors caused by spring-back and compare with another design method named 'force descriptor' [11] that focuses on evaluating the traction distributions on the metal sheet to design the mold shape. Compared to metal forming, glass thermoforming has additional error sources caused by viscous flow at high temperatures and shrinkage during cooling.

PGM was first attempted by the company Eastman Kodak in the 1970s [12]. Without numerical simulations, the early attempts based only on experimental trials failed because the random errors in mold fabrication defeated the efforts in error compensation to achieve optical precision. At the beginning of this century, several successful cases in PGM appeared, which were due to the help of finite element (FE) simulations where machining errors did not interfere. For example, Jain and Yi [13] developed a viscoelastic model of glass and demonstrated the feasibility of simulation-based mold design to achieve a specific lens. Through FE simulations and experiments, Wang et al. [14] demonstrated that the error compensation in mold design was effective in molding high-precision aspherical glass lenses. Zhou et al. [15] studied the viscoelasticity behavior of optical glass during the ultraprecision lens molding process, and they indicated that the creep and stress relaxation could be described based on Burgers model and Maxwell model. Yan et al. [16] investigated the effect of temperature distribution on PGM through thermo-mechanical FE simulations and pointed out that incomplete heating worsened the glass-forming performance.

Finite element (FE) simulation of a glass forming process generally takes tens of hours even for two-dimensional problems, therefore, it is inefficient to achieve error compensations through direct simulations, leaving alone the uncertainties in actual material and process parameters and mold machining errors when launching an FE-simulation-based mold design in actual production. Hence, the research on PGM was based on case studies, i.e., having a specific lens shape and exhibiting success in making it through simulation-assisted mold design and PGM [13, 14]. There is not a breakthrough in the design tool for PGM that can compensate for the form errors of glass to suit the fast update of curved glass products which happens in the cover glass industry (unlike lenses, the product cycle of cover glass is generally less than a year).

The basic requirement of a design tool for PGM is a surrogate model (SM) to predict error compensation profiles from the input of a glass geometry and process parameters. To this end, the machine-learning (ML) approaches, making predictions by learning a large set of well-defined data, are worth exploring. Recently, neural network (NN) models have been used in a wide range of engineering fields because of their excellent approximation performance and adaptability to multiple variables [17]. For example, Sivanaga et al. [18] proposed an NN model to predict the optimum process parameters of a wire-cut electric discharge machining. Choi et al. [19] established an NN model to predict the spring-back behavior in forming an electric-vehicle motor component based on geometric features and material properties. In addition, combining NN-based surrogate models with optimization algorithms can realize automatic parametric optimization. Tsai and Luo [20] combined NN and genetic algorithm (GA) to determine the optimal injection molding parameters to fabricate plastic lenses. They demonstrated that the model could help to meet the desired forming accuracy. While NN-based surrogate models have been used in multiple engineering fields, it is noted that there is not such an attempt in PGM in the open literature.

This work proposes a dimensionless backward propagation neural network (BPNN) model to replace time-consuming FE simulations or experiments in PGM. The model relates the inputs, i.e., geometric features and forming parameters, to the outputs, i.e., error compensation profiles for precision mold designs. First, we used the data set generated by FE simulations to train the BPNN. Then, we test the method based on industrial mold design data after trial-and-error cycles to meet the required accuracy of curved cover glass products. It is shown that the BPNN, after training, can be used to predict mold designs for molding 3D glass bodies with a tolerance of ~ 0.2 % of the maximum glass dimension. Therefore, with the validations using both FE simulations and industrial data, it is asserted that the proposed surrogate model can replace the costly simulations and experimental trials to quickly achieve precision mold designs.

# 2. Methods

### 2.1 Materials selection and properties determination

We aim to deform a piece of flat glass into a target 3D shape by molding with a set of upper and lower molds. Graphite is the mold material commonly used in the glass manufacturing industry due to the advantages such as low fabrication costs and nonstick to glass at high temperatures. It is also possible to use glassy carbon (GC) as the mold material with mechanical properties superior to graphite but difficult to machine [21]. To minimize the time cost of FE simulation, the molds in our simulations are set to be rigid so that the deformation of them will not consume computational resources. The only property of molds needed to be considered is thermal expansion which significantly influences the forming results of glass products. Hence, we set a thermal expansion to the rigid molds with the coefficients of thermal expansion (CTEs) referring to GC or graphite, as shown in Table 1 [22, 23].

Two types of glass material are considered in this study. One is aluminosilicate, represented by Corning's gorilla glass (GG); the other is borosilicate, referring to Schott's BK7 glass. Table. 1 [24] lists the mechanical and thermal properties of glass and mold materials used in our simulations.

**Table 1** The properties of glass, glassy carbon, and graphite.

| Material | Property | Value |
|---|---|---|
| Glass | Density ($g/cm^3$) | 2.5 |
| | Young's Modulus (GPa) | 76.7 (GG), 82 (BK7) |
| | Poisson's Ratio | 0.275 (GG), 0.206 (BK7) |
| | CTE ($10^{-6}$/°C) | 8.1 (GG), 8.3 (BK7) for $T < T_g$<br>12 (GG), 18.6 (BK7) for $T > T_g$ |
| Molds | CTE ($10^{-6}$/°C) | 2.5 (GC), 4.5 (Graphite) |

The viscoelasticity of glass in the time domain can be described by the Prony series (i.e., the generalized Maxwell model) that expresses the dimensionless relaxation modulus by using three variables: $g$, $k$, and $\tau$, representing the shear modulus ratio, bulk modulus ratio, and stress relaxation time at a reference temperature $T_0$, respectively. In this study, the bulk relaxation of glass is not considered (only happens in high-pressure experiments), so $k$ is set to be 0. The Prony series is shown below, and the parameters of the two kinds of glass are shown in Table 2 [24].

$$g(t) = 1 - \sum_{i=1}^{N} g_i (1 - \exp(-t / \tau_i)) \tag{1}$$

**Table 2** The parameters of the Prony series functions.

| Glass | $g$ | $k$ | $\tau$ |
|---|---|---|---|
| GG | 0.999 | 0 | 37.143 |
| BK7 | 0.999 | 0 | 0.00012 |

The effect of temperature on viscoelasticity can be described by the dependence of the instantaneous stress on temperature and by a reduced time concept. The shear stress influenced by temperature is written as:

$$\tau(t) = G_0(\theta) \int_0^t g(\xi(t) - \xi(s))\dot{\gamma}(s)ds \tag{2}$$

where the shear modulus $G_0$ is temperature dependent and $\xi(s)$ is the reduced time with defined as:

$$\xi(t) = \int_0^t (1/A(T(s)))ds \tag{3}$$

In Eq. (3), $A(T(s))$ is named shift function to reflect the temperature effect on the time scale (i.e., temperature-time superposition (TTS)). In this work, the Williams-Landel-Ferry (WLF) function is used as the shift function, given by:

$$\log_{10} A = -\frac{C_1(T - T_0)}{C_2 + (T - T_0)} \tag{4}$$

where $T_0$ is the reference temperature at which the relaxation data are given, and $C_1$ and $C_2$ are constants, respectively. Table 3 [24] shows the values of the parameters in the WLF function used in our simulations.

**Table 3** The parameters of the WLF function.

| Glass | $T_0$ (°C) | $C_1$ | $C_2$ |
|---|---|---|---|
| GG | 570 | 36.84842 | 1204.485 |
| BK7 | 685 | 5.01 | 179.4 |

## 2.2 Numerical models of glass and molds

3D glass covers have a wide variety of complex geometries, especially for VR/AR glasses. In this work, we start with the glass products having revolved profiles, which can be simplified as axisymmetric models in FE simulations. The outer profiles of these models are presumed to be aspherical or spherical functions, expressed as:

$$y_o(x) = \frac{cx^2}{1+\sqrt{1-(K+1)c^2x^2}} + a_1x^2 + a_2x^4 + a_3x^6 + \ldots \quad (5)$$

where $c$ is the surface curvature, $K$ is the conic constant, $a_i$ is the aspheric coefficient, $x$ is the radial coordinate and $y_o$ is the height coordinate. We assume that the thickness of a cover glass product is uniform; hence a cover glass product is represented by Eq. (5) and thickness $t$ and the inner profiles $y_i$ are obtained though CAD software. A flat glass blank to be deformed into a curved body must have the same volume as the final product. In our cases, the initial thickness of a flat blank shall be within 10% larger.

In a molding process, softened glass will fill the forming cavity during the molding step at the highest temperature, named the molding temperature, and then begin to shrink with decreasing temperature. Therefore, the dimension of the forming cavity shall be the same as that of the glass cover at the molding temperature. Due to the difference in coefficients of thermal expansion of glass and mold material (e.g., graphite), the dimensions of molds shall multiply a coefficient $m$ expressed as:

$$m = \frac{1+\alpha_{glass}\Delta T}{1+\alpha_{mold}\Delta T} \quad (6)$$

Where $\alpha$ is CTE with subscript refereeing to a specific material, and $\Delta T$ is the temperature difference between the molding temperature and room temperature.

### 2.3 Form error compensation

The forming cavity of molds can be established through the profile points of the generatrix. The sets of profile points of molds are labeled by their horizontal coordinates, given by $x$. Hence, the profile heights $y_{um}(x)$ and $y_{lm}(x)$ (corresponding to the upper mold and lower mold), being functions of $x$, are to be updated to achieve precision molds and their initial values, $y_{um}^0(x)$ and $y_{lm}^0(x)$, are based on the scaled glass profile $y(mx)$. After forming simulation with the initial version of molds, the obtained room-temperature glass profiles lead to another set of profile points, denoted by $y_{ig}(x)$ and $y_{og}(x)$, where the subscript $ig$ and $og$ denote the inner and outer surfaces of the molded glass. The surface profile deviations between the molded glass and designed glass can be used as error compensations to redesign the mold, and the error compensation profiles for upper and lower molds are then defined as $\Delta y_u(x) = y_i(x) - y_{ig}(x)$ and $\Delta y_l(x) = y_o(x) - y_{og}(x)$, respectively. Finally, these compensations shall be added to mold profiles of the last step ($i$) to get a new version ($i$+1) of molds until the precision molds that can form the glass profile with required accuracy are obtained, i.e., $y_{um}^{i+1}(x) = y_{um}^i(x) + \Delta y_u(x)$ and $y_{lm}^{i+1}(x) = y_{lm}^i(x) + \Delta y_l(x)$.

### 2.4 Neural network (NN) model

The profile deviations between initial molds and precision molds after error compensations are the key information in PGM mold design. This information, named form-error compensations (FECs), can be obtained through FE simulations or trial-and-error experiments. In this work, we test the hypothesis that a BPNN model can replace FE simulations to predict FECs for achieving precision molds.

A smooth and curved surface profile can be represented by a finite number of grid points with the local geometrical information to the second order derivatives, i.e., represented by inclination $angle(x) = \tan^{-1}(\frac{dy}{dx(x)})$), and curvature $\kappa(x) = \ddot{y}(x)/(1 + \dot{y}(x)^2)^{3/2}$. These geometrical features are critical for the precision mold design as they affect the local stress and strain states during a forming process, as the strains of a plate under bending can in principle be expressed as functions of inclination angle and curvature. As shown in Fig. 1, four geometrical features of glass product are selected as input variables: coordinate, thickness, Gaussian curvature (*K*), and inclination for a BPNN. The PGM parameters, in particular annealing rate and molding temperature, are also considered to be critical and set as input variables. Moreover, we use the maximum radial coordinate, $R_{max}$, of the revolved glass body, to nondimensionalize the geometrical features and FECs. $X = x/R_{max}$ is the dimensionless coordinate with the maximum value of 1; $T = t/R_{max}$ is the dimensionless thickness of the revolved glass body; $\overline{K} = K \times R_{max}^2$ represents the dimensionless Gaussian curvature; $\overline{FEC} = FEC/R_{max}$ represents the dimensionless of form-error compensation, which is the output of the BPNN.

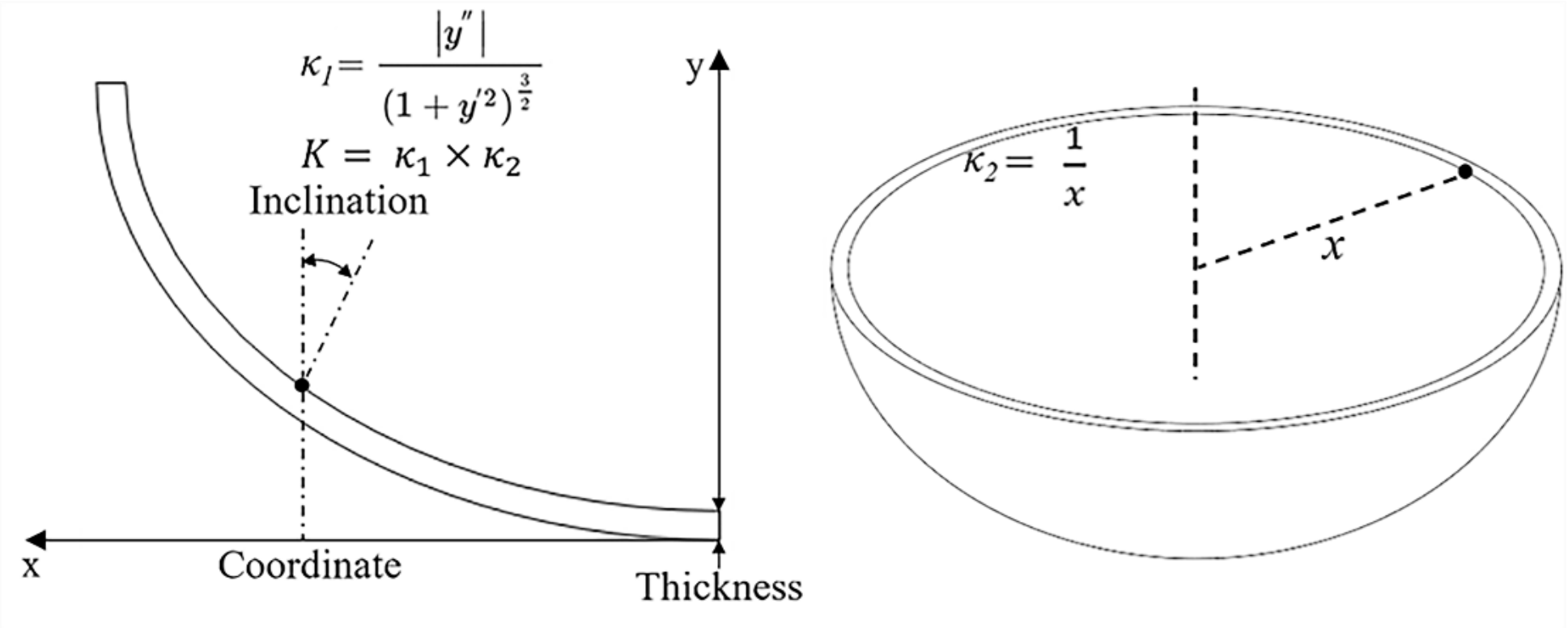

**Fig. 1** Geometric features used as input variables in the BPNN model.

# 3. Results and Discussion

## 3.1 Precision molds

We first demonstrate the results of error compensation based on the simulations with glassy carbon molds (i.e., assuming a small CTE of $2.5\times10^{-6}$ /°C) to achieve a pre-defined glass profile given by:

$$y(x)=\frac{0.04x^2}{1+\sqrt{1+0.04^2x^2}}+1.1\times10^{-5}x^4+3.9\times10^{-7}x^6+7.3\times10^{-10}x^8 .$$

The axisymmetric models of the glass and molds used in FE simulation are shown in Fig. 2(a), and the contact statuses between the glass and molds after the forming, annealing, and cooling stages are shown in Fig. 2(b-d). It can be observed that the glass does not contact mold surfaces when temperature decreases, causing thickness and profile deviations, which are collected to evaluate the forming performance, as shown in Fig. 3. With the molds designed according to the geometry of the glass cover, the forming accuracy is low. With a target thickness of 0.7 mm, the actual thickness of the molded glass varies with a maximum deviation of 4.0 μm. For the surface profiles, as shown in Fig. 3, the deviation of the inner surface gradually increases from the center area to the edge with a maximum deviation of 65 μm, and the deviation of the outer surface varies similarly with a maximum value of 60 μm. Such surface and thickness deviations cannot meet the accuracy requirements of precision forming of 3D glass, so FECs must be applied in mold designs.

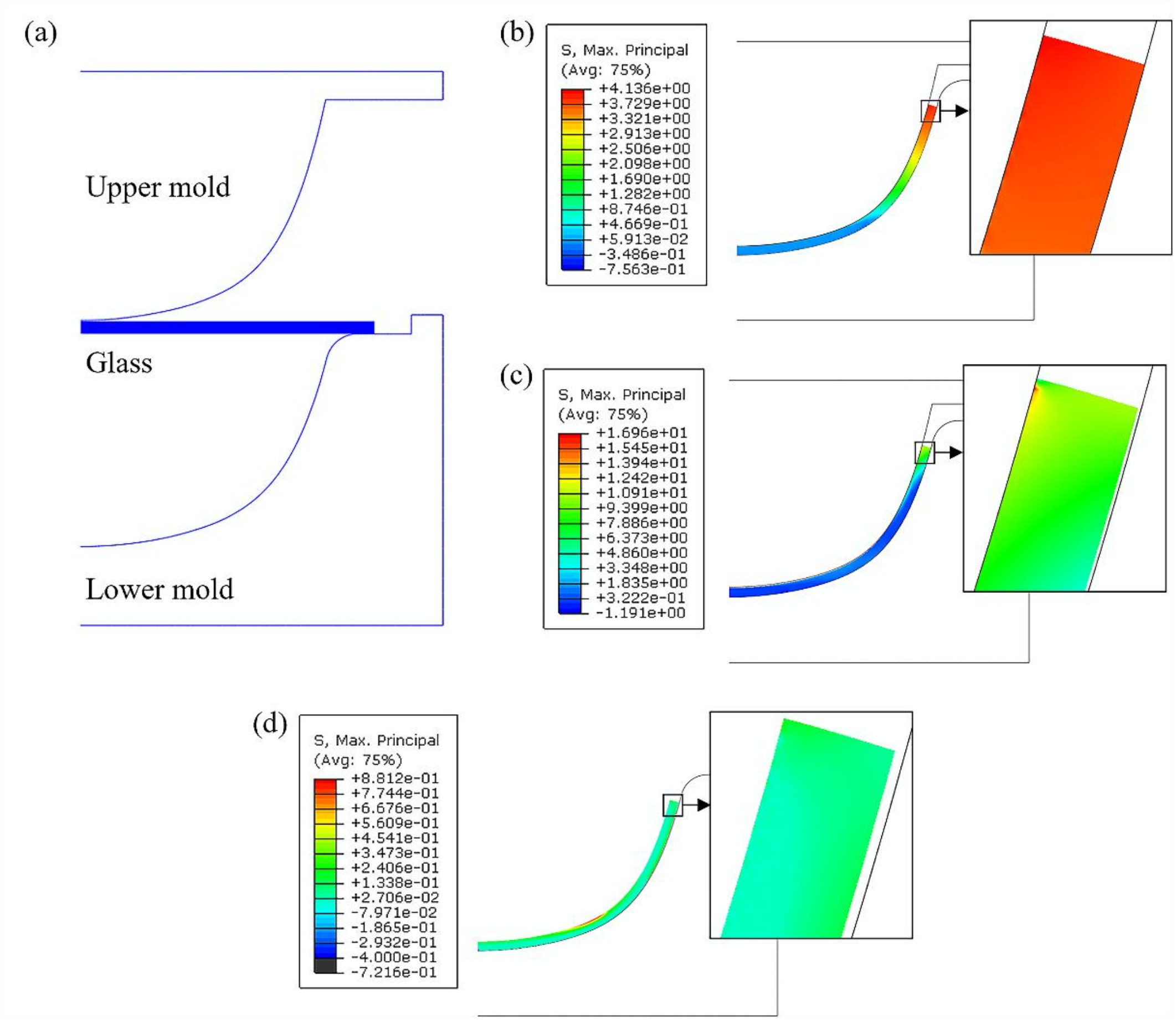

**Fig. 2** (a) axisymmetric models of glass and molds, (b) the contact status after forming stage, (c) the contact status after annealing stage, (d) the contact status after cooling stage.

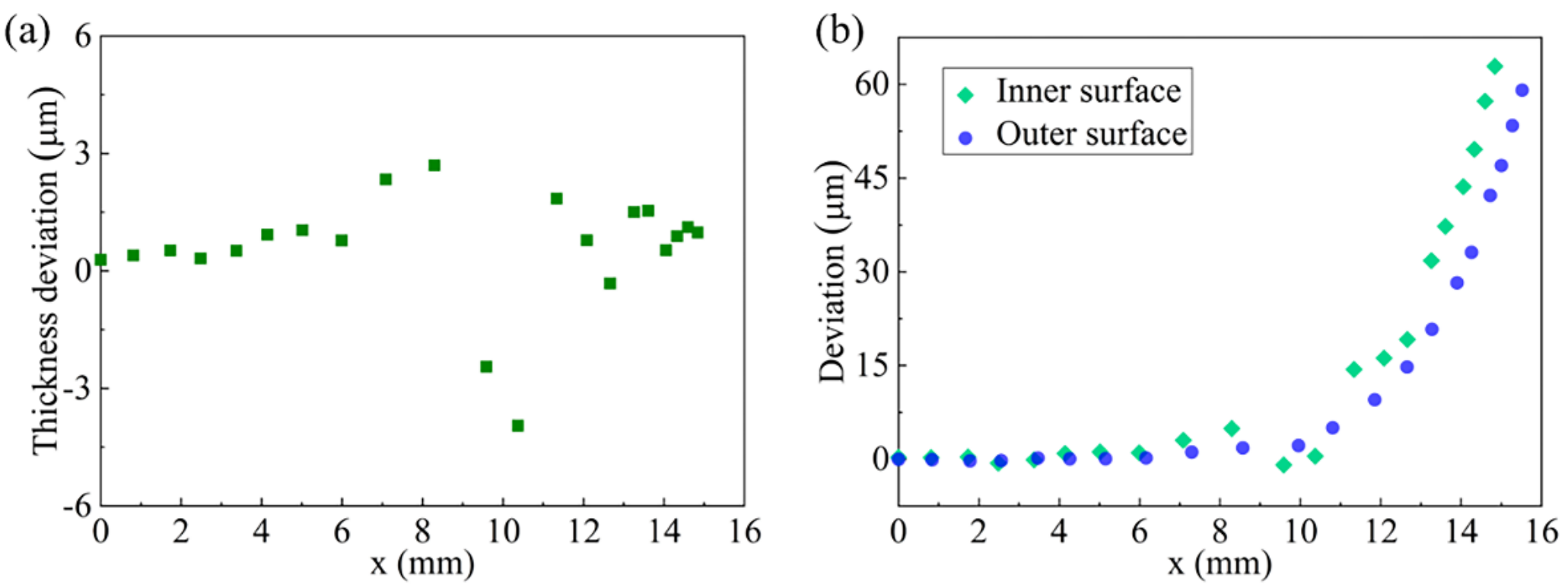

**Fig. 3** Simulation results before compensation: (a) thickness deviation, (b) surface deviation.

The approach described in section 2.3 is adopted to update the lower and upper molds, and the updated molds are used for the next simulation. Fig. 4 exhibits the forming performance after applying the FECs, the surface deviations can be reduced to below 2 μm and the thickness deviation is less than 3 μm, therefore, the one-step error compensation already achieves a good forming accuracy.

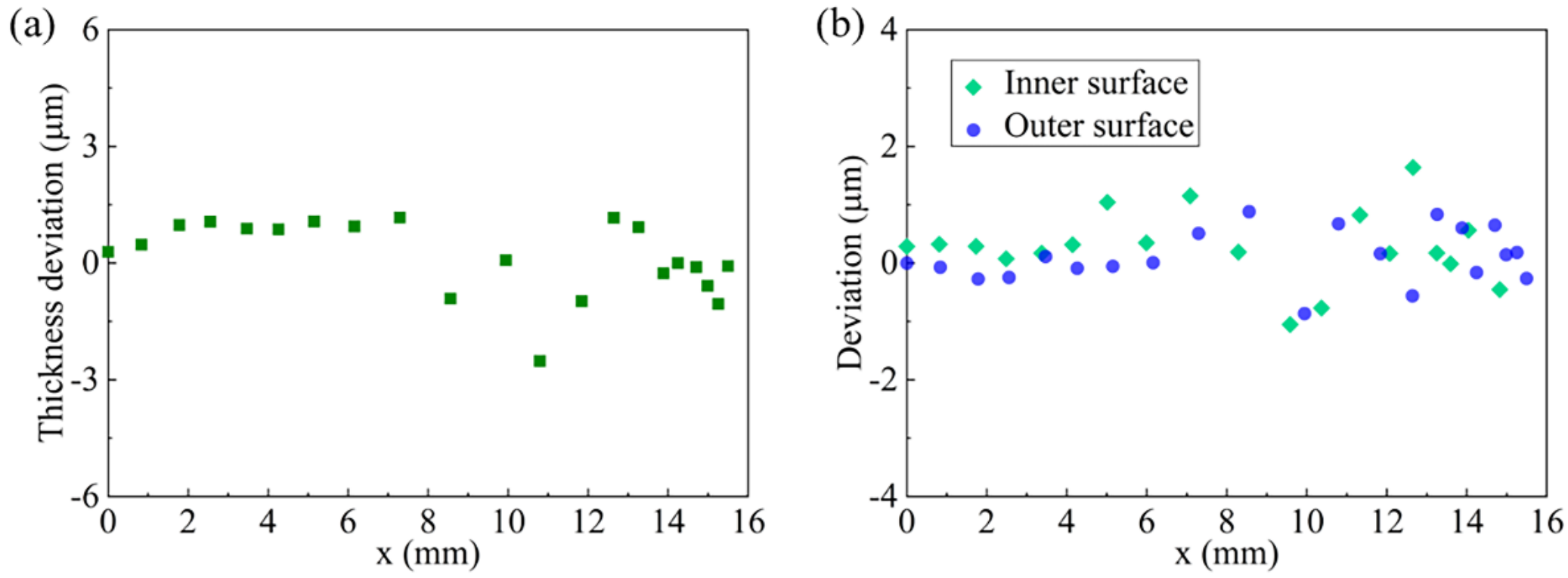

**Fig. 4** Simulation results after compensation: (a) thickness deviation, (b) surface deviation.

As graphite is more commonly adopted as mold material in the cover glass industry, we repeat the mold design process using the thermal properties of graphite. Fig. 5 shows both the surface deviation and thickness deviation of a glass cover formed by molds without compensation and after compensation. Before the molds are compensated, the maximum deviations in thickness, inner profile, and outer profile are 3.5, 28, and 22 μm, respectively. After the compensation, the forming accuracy has been significantly improved. The thickness deviations are reduced to less than 2 μm, and the deviations of the inner and outer surfaces are also below 2 μm. Therefore, it can be concluded that the error-compensation method is suitable for carbon-based molds with CTE between 2.5 – 4.5 $\times 10^{-6}$ /°C.

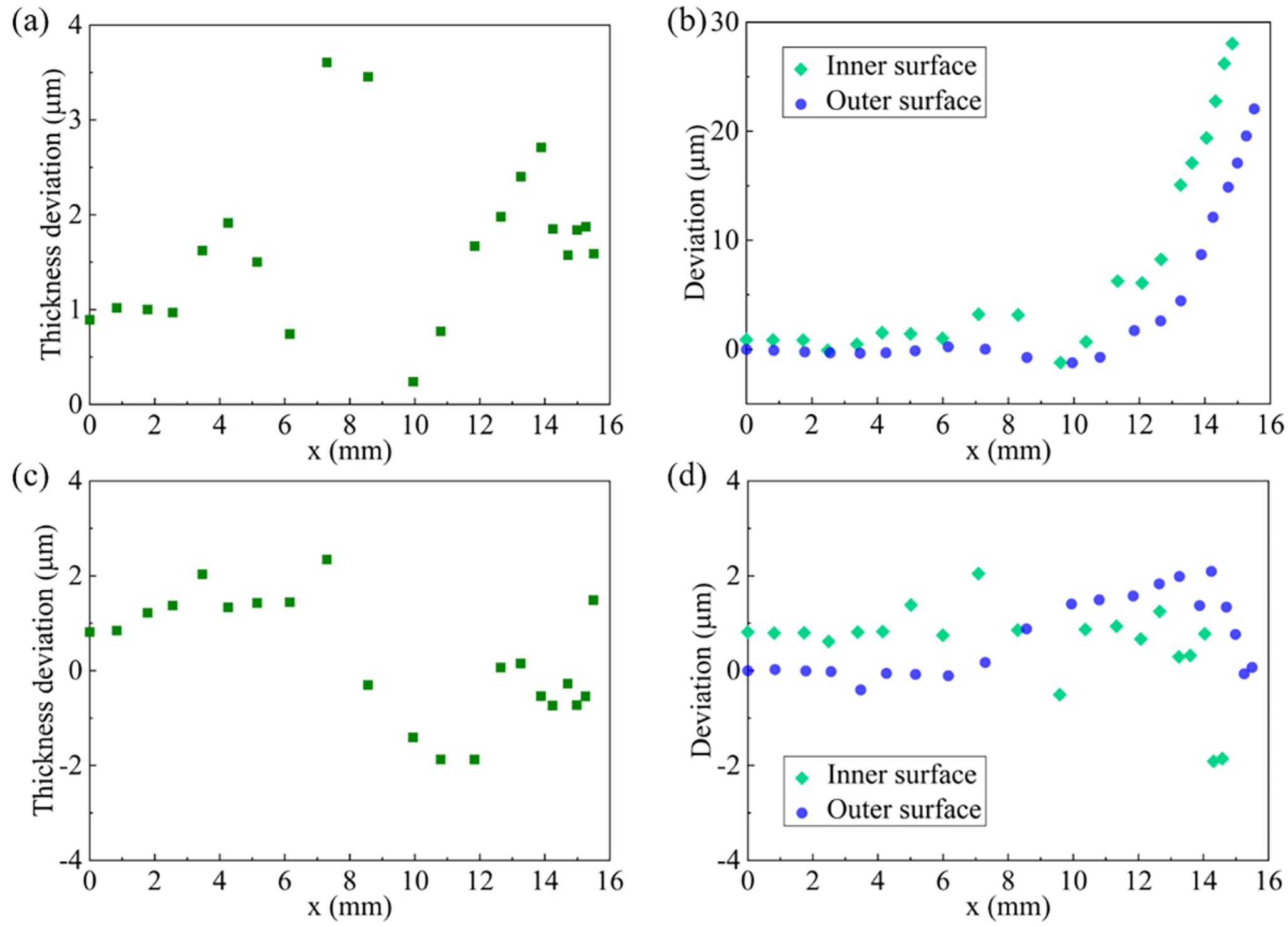

**Fig. 5** Simulation results using graphite molds. (a) thickness deviation before mold compensation, (b) surface deviation before mold compensation, (c) thickness deviation after mold compensation, (d) surface deviation after mold compensation.

### 3.2 Effect of machining errors on surface deviation

The above process is based on the premise that there are no fabrication errors on the mold surfaces, and the molds after the error compensation can be regarded as precision molds. However, in actual conditions, precision molds are impossible to obtain because machining errors will be generated during the mold manufacturing process. Therefore, it is necessary to study the effect of machining errors on the surface deviation of molded glass when precision molds, i.e., the FECs, have been determined through simulations.

We study the effect of machining errors by adding them to the precision molds introduced in the previous section. As shown in Fig. 6, a set of random numbers $y_e$ are added to the profile heights of molds $y_{um}(x)$ and $y_{lm}(x)$ to obtain the mold profiles with random errors $y_{um}^{ma}$ and $y_{lm}^{ma}$, where the superscript $ma$ represents the mold with machining errors. The effect of machining errors can be quantitatively expressed as an amplification factor $a_{ma} = \Delta y / y^{ma}$.

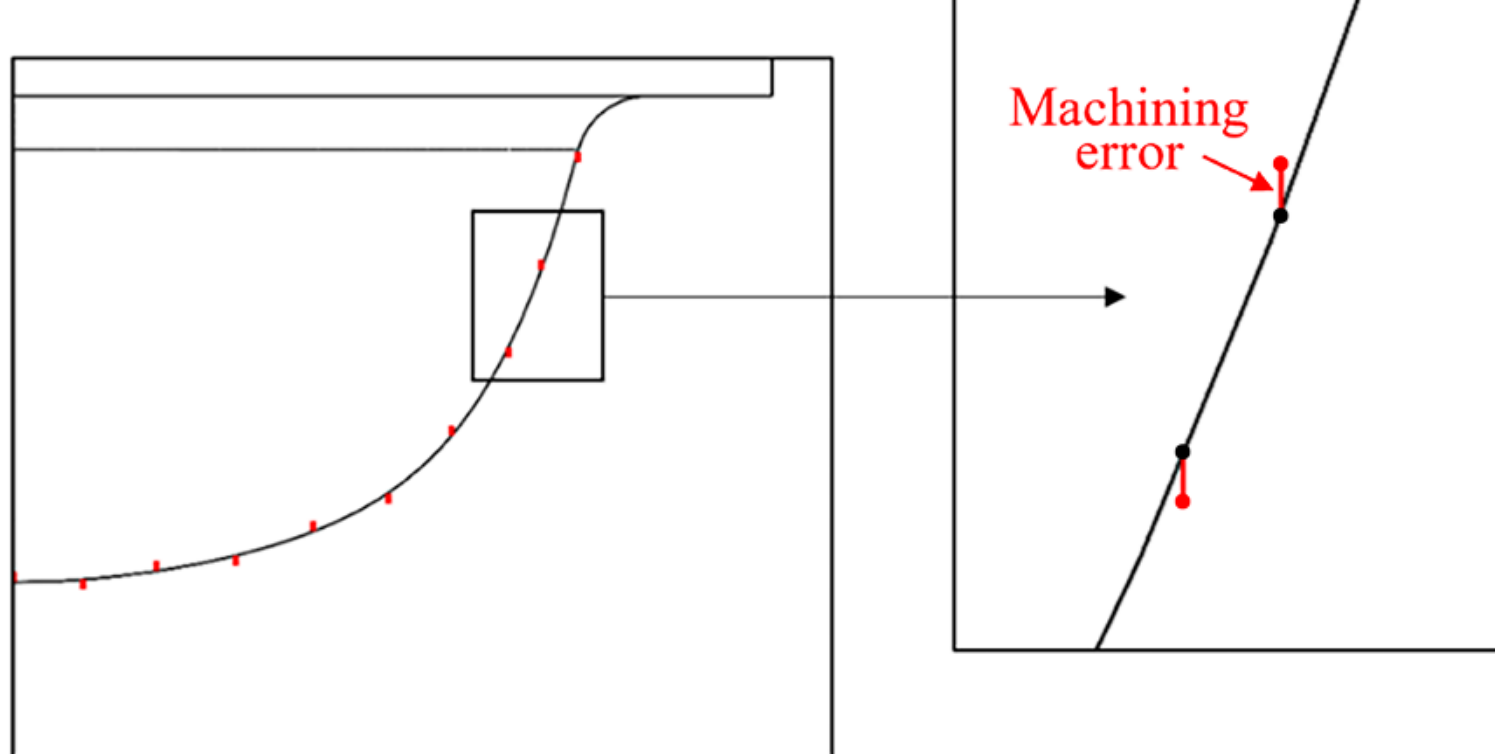

**Fig. 6** Schematic of adding machining errors to a precision mold profile.

We anticipate that the largest machining error must be below ±30 μm, which is achievable using common CNC machining systems. Therefore, we study the effect of three different levels of quality control, i.e., the tolerances of ±10 μm, ±20 μm, and ±30 μm, and focus on how the different levels of tolerance applied to the molds lead to different levels of forming errors in molded glass, i.e., the tolerance amplification factor from machining capability to forming accuracy. In this study, the glass surface deviations (below 2 μm) formed by precision can be neglected, because the surface deviations formed by the molds with machining errors become much larger than those. To generalize our work, many glass profiles are studied and herein we choose three of them (named Glass I, Glass II, and Glass III) to demonstrate and evaluate the tolerance amplification. The profiles of Glass I and Glass III are $y(x) = 12 - \sqrt{144 - \frac{9}{25}x^2}$ and $y(x) = \frac{8}{225}x^2$, while Glass II is the same as the glass cover discussed in Section 3.1.

Fig. 7 exhibits three machining error distributions and the surface deviations of glass covers formed by the molds with the corresponding machining errors. It is noticeable that the distribution of surface deviation on a glass surface is analogous to the corresponding machining error distribution. Fig. 7(b) exhibits the surface deviations of Glass I when the tolerance of machining errors is ±10 μm. At the point $x = 8$ mm, the inner surface deviation maximizes at 11.6 μm, which can be considered as an amplifying effect due to the machining error of 9.4 μm, leading to an amplification factor of 1.2. As shown in Fig. 7(c) and (d), the outer surface deviation of Glass II at point x = 2.6 mm is 27.1 μm, which is 1.4 times the corresponding machining error (18.7 μm) added to the lower mold. When the machining error added to the upper mold at point $x = 12.3$ mm is 16.1 μm, the inner surface deviation of Glass II is 27.9 μm, which is 1.7 times the machining error. Fig. 7(e) and (f) show the distribution of machining errors with a tolerance of ±30 μm and the forming performance of Glass III. At point $x = 2.9$ mm on the upper mold, the machining error added on is 26.6 μm; at a similar location on the glass inner surface, the surface deviation is about 40.8 μm, ~ 1.53 times the machining error. Similarly, a machining error of about 18.5 μm is added to the lower mold profile at point $x = 9.7$ mm, leading to a deviation of 29.2 μm on the outer glass surface, ~1.57 times the machining error.

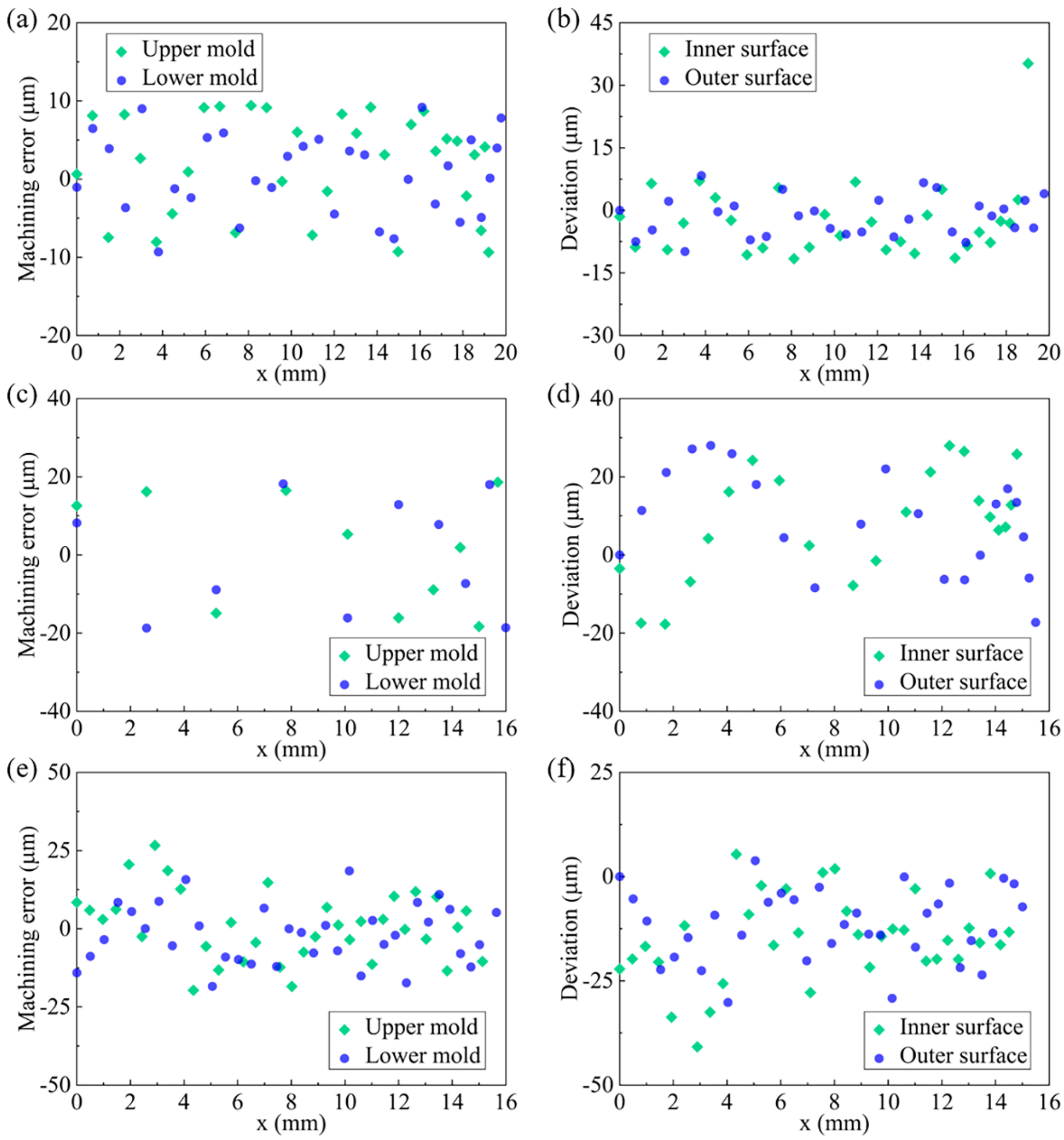

**Fig. 7** (a) Machining errors distribution with a tolerance of ±10 μm, (b) surface deviation of Glass I when machining errors tolerance is ±10 μm, (c) machining errors distribution with a tolerance of ±20 μm, (d) surface deviation of Glass II when machining errors tolerance is ±20 μm, (e) machining errors distribution with a tolerance of ±30 μm, (f) surface deviation of Glass III when machining errors tolerance is ±30 μm.

The above results indicate that the amplification of error tolerance from mold machining to glass thermoform is less than a factor of 2 based on the knowledge of precision molds. Note that when the tolerance of machining errors is ±30 μm, the glass form errors can reach 50 μm, which is almost the limit of 3D cover glass products in the industry. Hence, it is expected that when the precision molds or the FECs are unknown, it is very difficult to control the form errors for 3D cover glass manufacturing. In this case, a predictive model to provide FECs for precision mold design is of great significance for the industry.

### 3.3 NN model for FEC prediction

#### 3.3.1 NN model based on simulation data

A BPNN model with eight hidden layers is established to predict the FECs. Fig. 8 shows the structure of the BPNN. There are six neurons in the input layer and one neuron in the output layer. The numbers of neurons in eight hidden layers are 12, 12, 12, 12, 10, 10, 8, and 8 respectively. The ReLU function is chosen as the activation function in the model, while the mean square error is selected as the loss function to evaluate the accuracy of the model. Adam optimizer is chosen due to its high efficiency and strong adaptability.

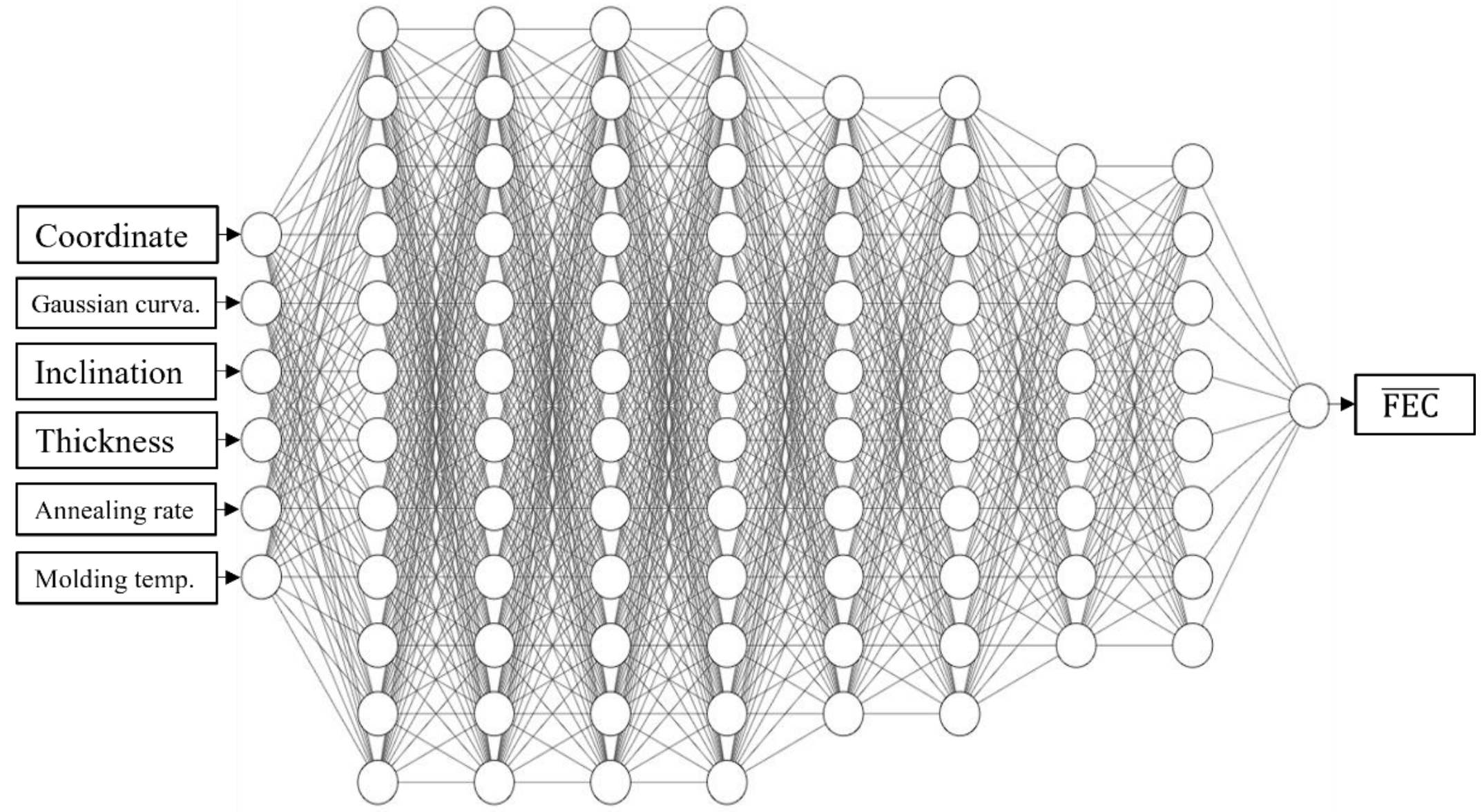

**Fig. 8** Structure of the BPNN based on simulation data.

The training dataset is obtained based on several virtual mold compensation processes with varied forming conditions and 3D glass profiles through FE simulations. 280 data sets are used to train and test the model with a ratio of 7:3, and $R_{max}$ in the database is smaller than 20. Note that the number of data sets is small compared to current data-driven methods. However, it is commensurate with other efforts in the manufacturing field to develop SMs replacing FE simulations and trial-and-error experiments (e.g., [18]), and our aim here is to test if the geometric features extracted from simulation-based mold designs, though limited in the size of datasets, are sensible to develop a SM with a reasonable capability of predicting FECs for PGM molds.

Fig. 9 exhibits the results of training in terms of the comparisons between the BPNN predictions and non-dimensional training and testing data (denoted by $\overline{FEC}$ in the figures). For the training group, the loss value is 1.94×10$^{-3}$. $R^2$ value, commonly used to describe the performance of a NN model, is close to 1 for the training group. For the prediction of data in the testing group, the loss value is 4.39×10$^{-3}$, and $R^2 = 0.92$. The results exhibited in Fig. 9 demonstrate that the BPNN has been adequately trained with a dimensionless prediction tolerance less than 0.01%.

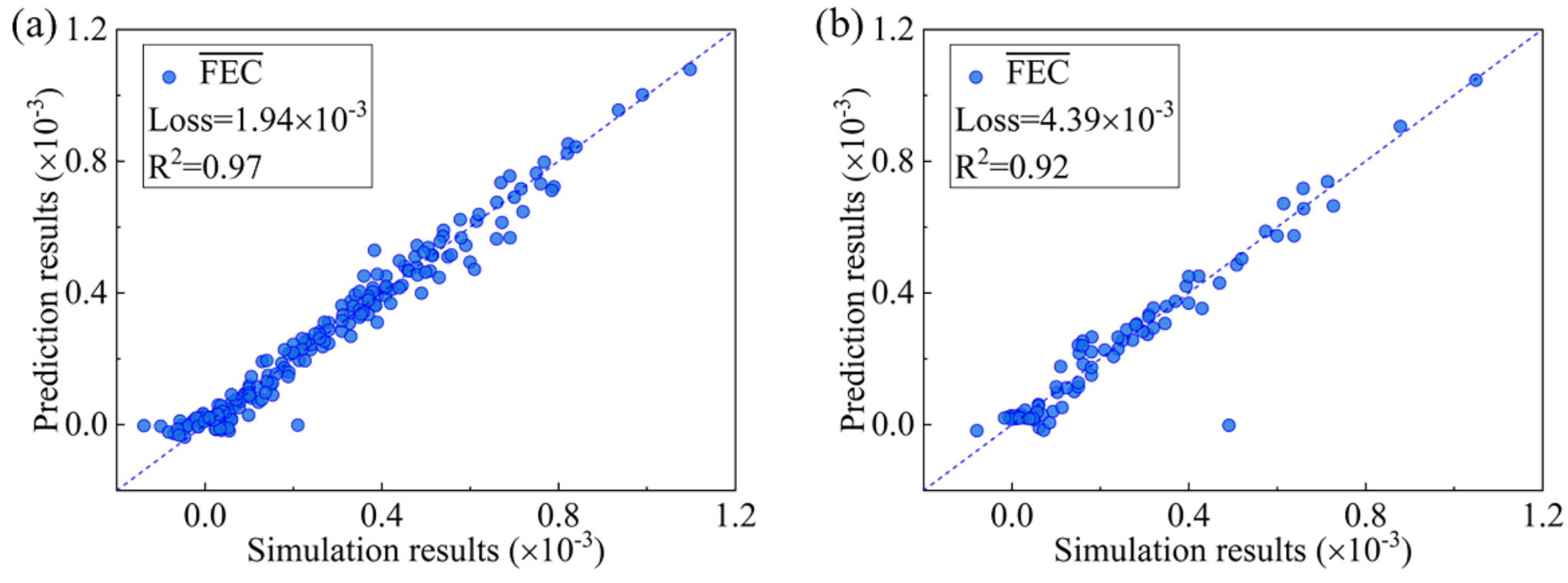

**Fig. 9** Prediction results of the BPNN: (a) training group, (b) testing group.

The applicability of this BPNN depends on whether it can assist the design of precision molds for thermoforming thin glass covers with different geometries. We first demonstrate its predictions of FECs for two glass profiles different from those in the training and testing datasets. FE Simulations to form these two glass covers were also conducted to determine the actual FECs for precision molds. The comparisons are shown in Fig. 10, demonstrating a good consistency with the loss values and $R^2$ similar to those shown in Fig. 9(a). It is noted that the differences between the two methods are less than 0.01% of $R_{max}$, much smaller than machining errors (note that our $R_{max}$ = 20 mm, hence the deviation is less than 2 μm). Hence, this BPNN model can predict the FECs needed to form revolved cover glass products with different profiles.

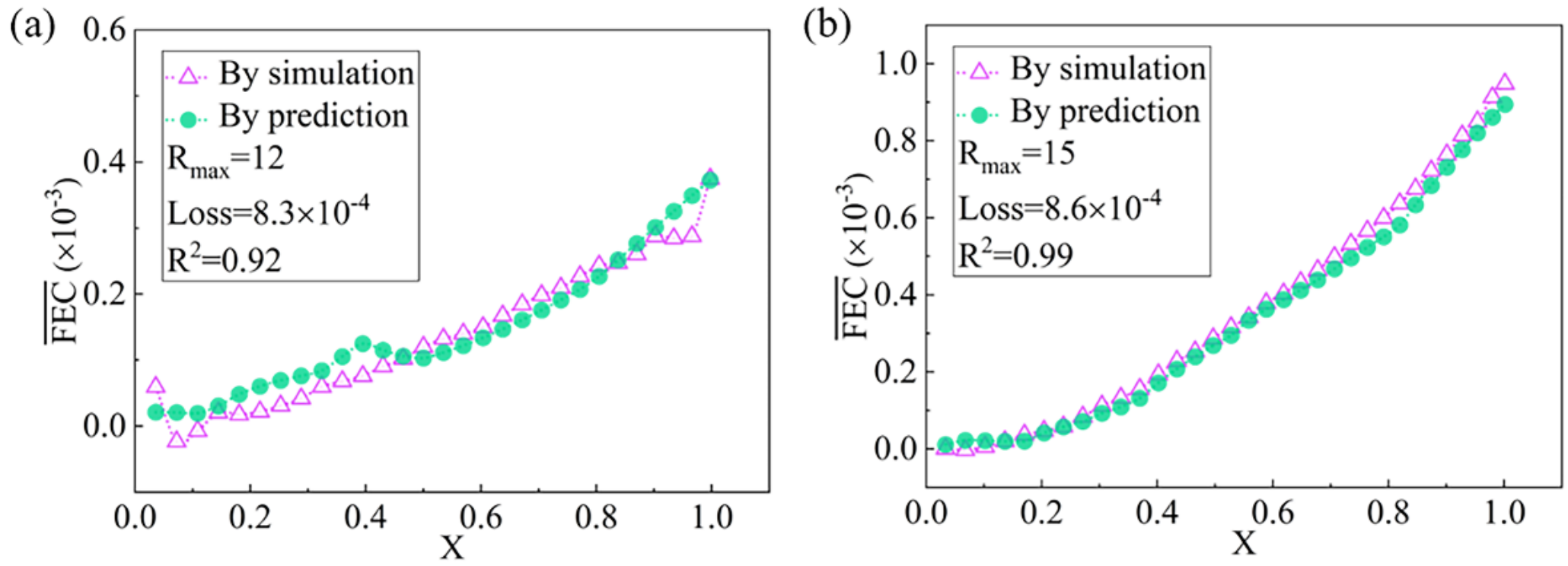

**Fig. 10** Prediction results of BPNN model: (a) validation group for glass with profile I, (b) validation group for glass with profile II.

After validating the BPNN model, we further test its prediction performance when the glass dimensions (i.e., the range of *x* coordinate) are much larger than those in the training set. This is to check whether the input of dimensionless geometric features is adequate to determine FECs. Considering that the maximum $R_{max}$ in the database is 20 mm, we demonstrate two revolved glass covers with larger radii of 40 mm and 68 mm. The prediction performance of the BPNN is still satisfactory in these two cases, as shown in Fig. 11. The prediction errors of FEC are mostly within 0.01%, except for the predictions at the edges of two glass geometries. The results shown in Fig. 11 indicate that the nondimensionalized BPNN can mitigate the impact of glass dimension on the prediction, allowing the BPNN trained by data of small glass covers to predict the FECs required for large glass covers.

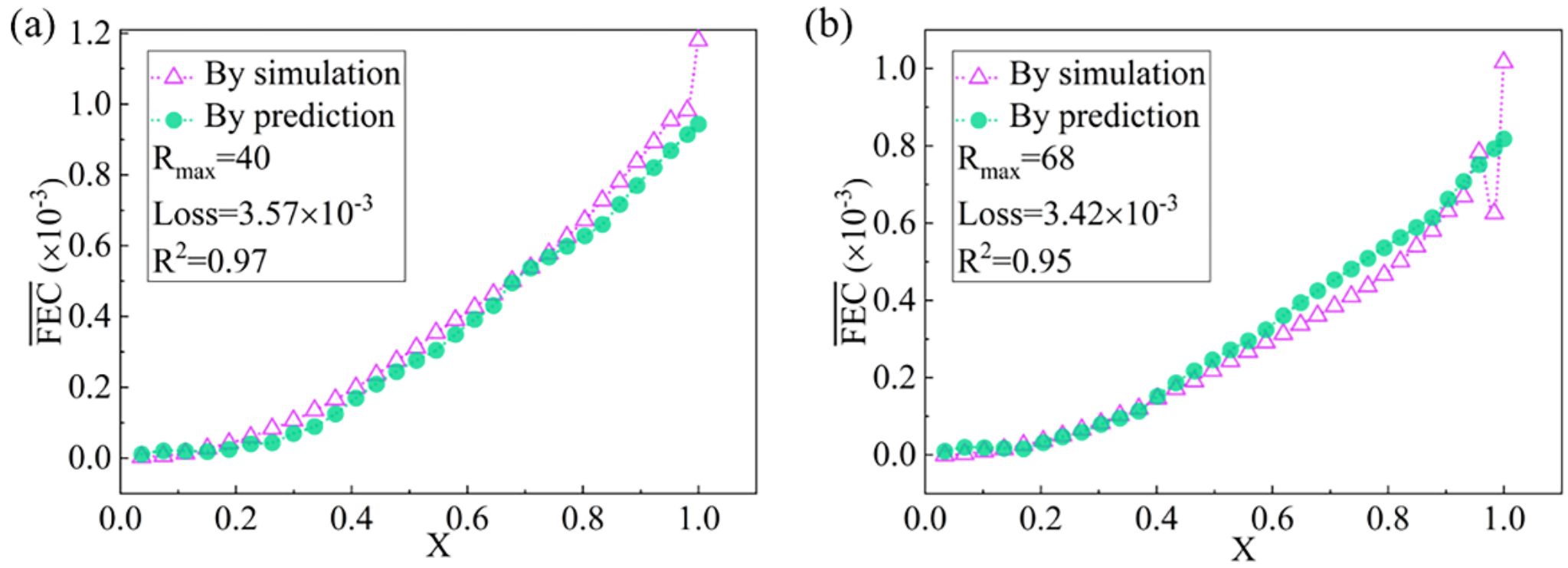

**Fig. 11** Prediction results of the BPNN for large glass profiles: (a) radius of 40 mm and (b) radius of 68 mm.

### 3.3.2 NN model based on industrial data

The data used to develop a NN-based SM in the previous section is idealized and greatly simplified. It is necessary to test whether the same method can be extended to industrial data, i.e., whether a NN model based on the proposed geometrical inputs of glass designs can predict the FECs in precision mold designs. In this section, smartphone cover glass molding data with different geometries and dimensions provided by the company (Biel Crystal Manufactory (H.K.) limited) are adopted to train a NN model. The cover glass designs are no longer axisymmetric, hence, the geometric data are labelled by their in-plane coordinates (x, y), which are normalized by the maximum in-plane size of each glass design. The NN is re-designed because the glass geometries are more complicated, as shown in Fig. 12. There are four neurons in the input layer (Different from Fig. 8, the inputs of process parameters are omitted because the company adopted the same process parameters), one neuron in the output layer for $\overline{FEC}$ and nine hidden layers. The numbers of neurons in the nine hidden layers are 13, 12, 12, 12, 6, 4, 4, 4, and 4, respectively.

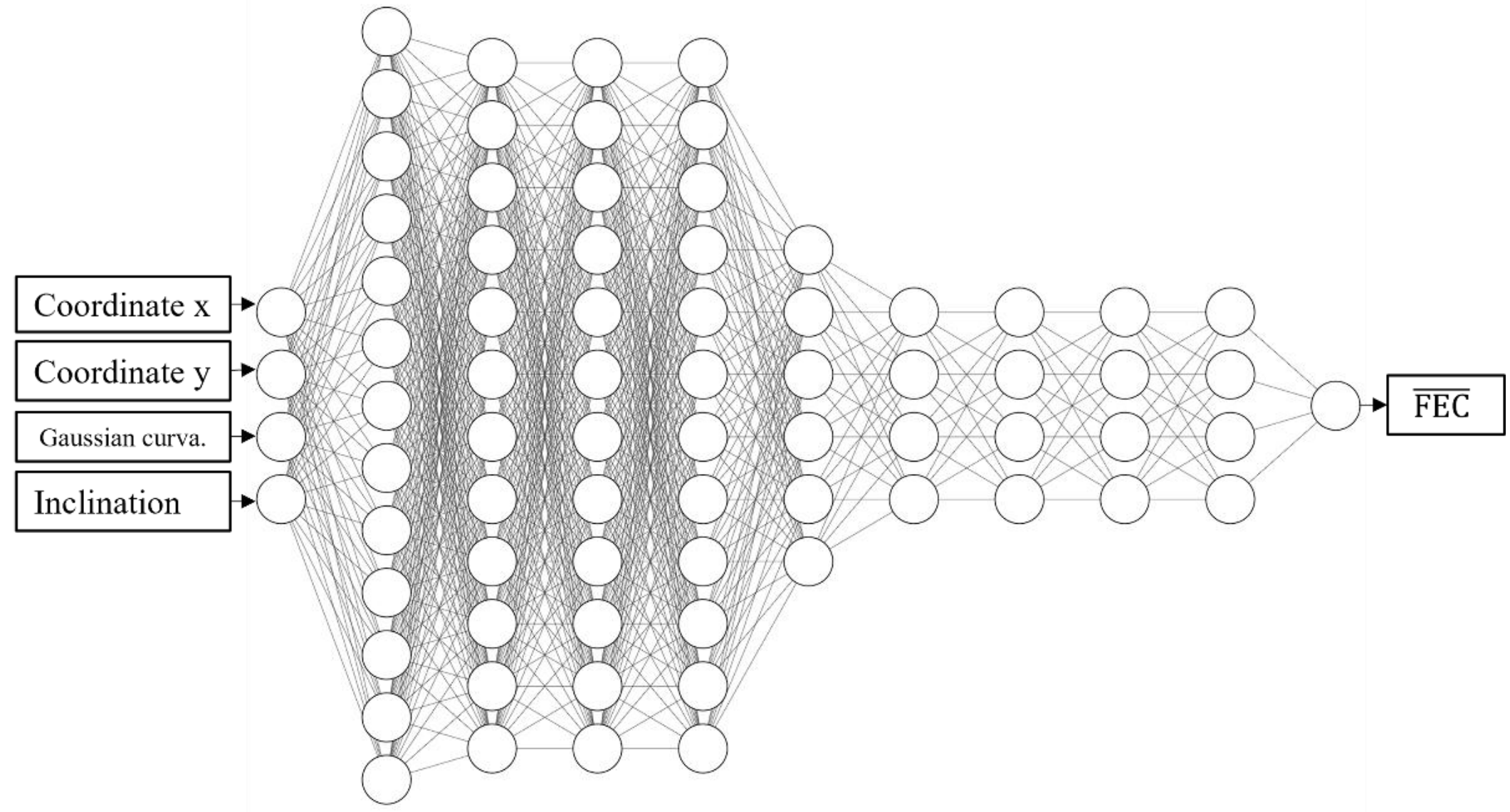

**Fig. 12** Structure of the BPNN based on experimental data.

The industrial mold designs were obtained after tens of trial-and-error cycles to update FECs. Unlike simulations, these mold designs were determined under the condition that mold fabrication inevitably brought about random errors to the designed profiles. Therefore, these mold designs might be biased to correct fabrication errors which do not have a zero mean and the accuracy of molded glass must be compromised with the tolerance of former errors being tens of micrometers.

By discretizing 4 sets of mold designs adopted in production, we obtain 2473 sets of geometrical data for training and testing and 618 sets of data for validation. Fig. 13 exhibits an example of the cover glass model and the comparisons between prediction results of the NN model and the industrial data, showing a satisfactory training result with $R^2$ = 0.96. The differences between the predicted and actual FECs are within 0.05% of the maximum glass dimension. But for the testing group and validation group, the prediction results are not as accurate as those for the training group, with $R^2$ values of 0.84 and 0.83 respectively. The maximum prediction errors of testing and validation groups are also larger than that of the training group with a value of 0.2%. While the predictions based on the proposed NN trained using industrial data have larger errors than those based on simulation data, it must be noted that the industrial model design process is affected by many uncertainties, such as machining errors, personal habits of form error compensation, glass properties, and processing parameters. Hence, we argue that the proposed data-driven approach can assist mold design for curved glass production.

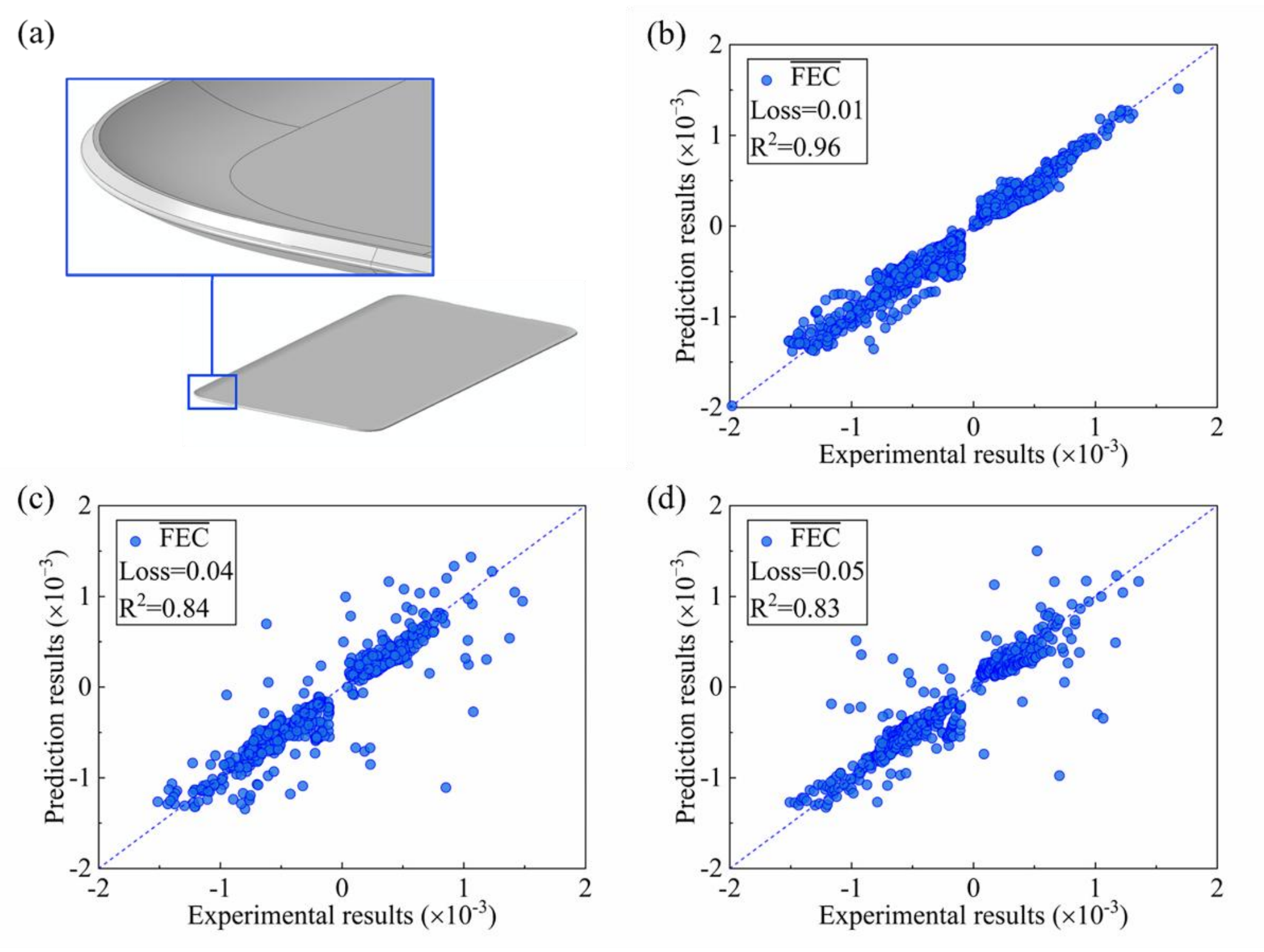

**Fig. 13** (a) An example of the smartphone cover glass model, (b) training group results, (c) testing group results, (d) validation group results.

# 4. Conclusions

Based on the simulations of the PGM process for shaping revolved glass profiles, the form errors can be compensated onto mold surfaces to achieve precision molds with a forming accuracy within 2 μm. With precision molds, the amplification factor from mold machining errors to glass forming errors is less than 2, which can be used to regulate the tolerance of mold machining errors in the industry of cover glass manufacturing. Designing precision molds requires the determination of FECs. Though they can be obtained through FE simulations or trial-and-error experiments (due to the existence of machining errors), these processes are very time-consuming and costly, sometimes impractical for complicated geometries. Mold designers may need faster tools to determine FECs even with a sacrifice of some accuracy; hence, the main contribution of this work is a dimensionless BPNN model which acts as a surrogate model replacing FE simulations. The BPNN model established in this work is proved to have a good performance in predicting the FECs based on the inputs of geometric features (inclination angles and curvatures) and thermoforming parameters. In the tested cases, the difference of FECs between the BPNN prediction and FE simulation is less than 0.01% of the maximum radial dimension of revolved glass covers. Because of the nondimensionalization, this BPNN, trained using data generated based on small glass profiles, can predict the FECs needed for glass profiles with much larger dimensions. Based on the industrial data, it is demonstrated that a BPNN model with the proposed geometrical inputs can also assist mold designs by predicting the FECs in the same accuracy level of the contemporary industrial practice.

## Statements and Declarations

**Competing Interests:** The authors declare that they have no known competing financial interests or personal relationships that could have appeared to influence the work reported in this paper.

## Data Availability

Data will be made available on request.

## Author Contribution

Yuzhou Zhang: Writing – original draft, Conceptualization, Methodology, Formal Analysis, Software. Mohan Hua: Methodology, Formal Analysis, Software. Jinan Liu: Resources. Haihui Ruan: Conceptualization, Supervision, Writing – review & editing.

## Acknowledgement

We gratefully acknowledge the financial support provided by the Hong Kong GRF (Grant No. 15210622) and by the industry (HKPolyU Project ID: P0039303).